\documentclass{aastex62}

\usepackage{amsmath}
\usepackage{units}

\begin{document}

\title{Blue straggler stars beyond the Milky Way. II. A binary origin
  for blue straggler stars in Magellanic Cloud clusters}

\author{Weijia Sun} 
\affiliation{Kavli Institute for Astronomy \& Astrophysics and
  Department of Astronomy, Peking University, Yi He Yuan Lu 5, Hai
  Dian District, Beijing 100871, China}

\author{Chengyuan Li}
\affiliation{Department of Physics and Astronomy, Macquarie
  University, Balaclava Road, Sydney, NSW 2109, Australia}
\affiliation{Department of Astronomy, China West Normal University,
  Nanchong 637002, China}

\author{Richard de Grijs}
\affiliation{Department of Physics and Astronomy, Macquarie
  University, Balaclava Road, Sydney, NSW 2109, Australia}
\affiliation{International Space Science Institute--Beijing, 1
  Nanertiao, Zhongguancun, Beijing 100190, China}

\author{Licai Deng}
\affiliation{Key Laboratory for Optical Astronomy, National
  Astronomical Observatories, Chinese Academy of Sciences, 20A Datun
  Road, Chaoyang District, Beijing 100012, China}
\affiliation{School of Astronomy and Space Science, University of the
  Chinese Academy of Sciences, Huairou 101408, China}
\affiliation{Department of Astronomy, China West Normal University,
  Nanchong 637002, China}

\begin{abstract}
We have analyzed populations of blue straggler stars (BSSs) in 24
Magellanic Cloud star clusters using multi-passband \textit{Hubble
  Space Telescope} images. We compiled a homogeneous BSS database,
containing both traditional and evolved BSSs. We uncovered a
sub-linear correlation between the number of BSSs in the cluster cores
and the clusters' core masses, characterized by a power-law index of
$0.51\pm 0.07$. For low stellar collision rates, the mass-normalized
number of BSSs depends only weakly (or perhaps not at all) on the
collision rate, implying that the binary-driven BSS formation channel
dominates. Comparison with simulations suggests that stellar
collisions contribute less than 20\% to the total number of BSSs
formed. Further tests, including analysis of the BSS specific
frequencies and their population numbers at larger cluster radii,
suggest that binary interactions may be their main formation channel,
hinting at an anti-correlation between a cluster's binary fraction and
its core mass.
\end{abstract}


\section{Introduction \label{sec:intro}}

Blue straggler stars (BSSs) are brighter and bluer than the
main-sequence turnoff (MSTO) in a given star cluster, yet they occupy
the extrapolation of a cluster's main sequence as defined by its bulk
stellar population. Since they were first discovered in the globular
cluster (GC) M3 by \citet{1953AJ.....58...61S}, BSSs have been found
in all known GCs \citep{2004ApJ...604L.109P}, as well as in open
clusters \citep{2008MNRAS.384.1263C} and dwarf galaxies
\citep{2007MNRAS.380.1127M}.

The prevailing scenarios for the formation of BSSs involve direct
collisions of single stars and/or binary systems
\citep{1964MNRAS.128..147M, 2003ApJ...588..464F} or mass transfer
and/or the coalescence of primordial binaries
\citep{1984ApJ...284..719I, 2009Natur.457..288K,
  2013MNRAS.428..897L}. We must examine the entire population of BSSs
in clusters in order to distinguish between the different formation
mechanisms. The first attempt to do so was made by
\citet{2003ApJ...588..464F}, who compared the predicted number of
collisional BSSs with actual observations in six GCs and found good
agreement between both numbers for most clusters, except for the
lowest-density GC. However, \citet{2004ApJ...604L.109P} detected a
weak anti-correlation of the numbers of BSSs with the theoretical
collision rate (see below), thus indicating that collisions cannot be
the main source of the observed populations of BSSs.

\citet{2009Natur.457..288K} found a robust correlation with an
intrinsic cluster property for their sample of 57 Galactic GCs: the
number of BSSs in a cluster's core exhibits a `sub-linear' correlation
with the cluster's core mass, $N_\mathrm{BSS,core}\propto
M_\mathrm{c}^{0.5}$. This can be understood as the result of binary
interactions if the binary fraction also depends on the core mass,
i.e., $f_\mathrm{bin}\propto M_\mathrm{c}^{-0.5}$
\citep{2012A&A...540A..16M}. However, \citet{2013MNRAS.428..897L}
found that including the numbers of binary stars in the cores of 30
additional Galactic GCs did not strengthen the correlation; the
resulting degradation of the correlation further complicated the
underlying issue as regards the origin of BSSs.

Indeed, the distinction between both formation channels is not as
clear-cut in real environments. On the one hand, a cluster's core
binary fraction is strongly regulated by dynamical interactions
\citep{2009ApJ...707.1533F}, thus directly influencing the formation
rate of BSSs. On the other, under some conditions, collisions are more
probable to occur between binary stars because of their large cross
sections \citep{2011MNRAS.416.1410L}. Increasing lines of evidence
suggest that both formation channels are closely
linked. \citet{2006MNRAS.373..361M} argued that the different origins
of BSSs may be related to their spatial distribution: BSSs in a
cluster's periphery then form through mass transfer, whereas BSSs
close to the cluster core most likely have a collisional
origin. Another piece of evidence came from the two distinct sequences
of BSSs observed in the color--magnitude diagram (CMD) of M30
\citep{2009Natur.462.1028F}.

To provide a better understanding of the prevailing BSS formation
mechanisms, we need to expand our samples to include younger and more
distant objects. Thus far, only a few studies have focused on
extragalactic GCs \citep{2013MNRAS.436.1497L,
  2018MNRAS.476.5274L}. The Magellanic Clouds (MCs) host numerous
clusters spanning a broad range of ages
\citep{2005ApJS..161..304M}. C. Li et al. (submitted) discovered a
branch of evolved stars which are much brighter than the normal
evolved stars in their host clusters. The most straightforward
interpretation of these stars is that they are all younger than the
cluster's bulk stellar population. Based on isochrone fitting, they
found that the masses of these young evolved stars are generally lower
than twice the average mass of their normal counterparts. Thus, Li et
al. attributed the origin of this younger population to evolved
BSSs---BSSs having evolved off the MS---which had already been
identified in the Galactic GCs M3, M13 \citep{1997ApJ...484L.145F},
and M80 \citep{1999ApJ...522..983F}. \citet{2016ApJ...830..139P}
reported a total number of evolved BSSs in the Galactic GC 47 Tucanae
(47 Tuc) that was comparable to the number of BSSs observed, leading
them to suggest a much shorter BSS lifetime than previously thought.

In this paper, we analyze 24 clusters in the MCs. We uncover a
sub-linear correlation between the number of BSSs in the cluster cores
and the clusters' core masses, with a power-law index of $0.49\pm
0.09$. The mass-normalized number of BSSs is only weakly dependent on
the mass-normalized collision parameter and, for higher collision
rates, it decreases with increasing collision parameter, clearly
indicating the predominance of binary disruption. Therefore we
conclude that the formation of the observed BSSs is likely dominated
by binary mechanisms, at least in most of our sample clusters.

This article is organized as follows. In Section \ref{sec:data} we
present the observations and data reduction procedures
applied. Section \ref{sec:result} reports our main results, including
any correlations and anti-correlations with cluster properties. Next,
we discuss the possible BSS formation channels in Section
\ref{sec:discussion}. Finally, we summarize our study in Section
\ref{sec:conclusion}.

\section{Observations and Data Reduction \label{sec:data}}
\subsection{Photometry}

The clusters in the present paper were selected from the {\sl Hubble
  Space Telescope} ({\sl HST}) Data Archive at the Space Telescope
Science Institute. We collected observational data obtained with the
Advanced Camera for Surveys/Wide Field Channel (ACS/WFC) and the Wide
Field Camera 3/Ultraviolet and Visible channel (WFC3/UVIS). Most
cluster data sets included a nearby `reference' field, which we used
to statistically estimate the contribution from the background/field
stellar population. For clusters without a reference field, we only
selected those objectw with sizes that were sufficiently small that an
edge of the science image could be used for field correction purposes
and to derive reliable structural
parameters. \citet{2017ApJ...839...64R} argued that UV-based surveys
are more efficient in identifying BSSs than optical surveys. However,
although for some clusters ultraviolet (UV) observations were
available (e.g., {\sl HST} program GO-14164; PI: A. Sarajedini), we
nevertheless only based our selection on the presence of data in
optical passbands. This was done because only very few clusters were
observed at UV wavelengths and we prefer to establish a homogeneous
observational basis for our work. We selected all intermediate-age
(\unit[1--3]{Gyr}-old) and old ($\sim\unit[10]{Gyr}$-old) clusters in
the MCs for which suitable {\sl HST} observations were available (for
more details, see Section \ref{sec:param}). Young massive clusters
were excluded, because these clusters do not exhibit a clear MSTO
point, which renders the selection of BSSs unclear. Detailed
information about our 24 sample clusters is presented in Table
\ref{tab:obs}.

\startlongtable
\begin{deluxetable*}{lllClll}
\tablecaption{Summary of the {\sl HST} observations of our Magellanic
  Cloud cluster (MCC) sample. \label{tab:obs}}
\tablewidth{700pt}
\tabletypesize{\scriptsize}
\tablehead{
\colhead{Cluster} & 
\colhead{Galaxy} &
\colhead{Camera}  & 
\colhead{Exposure Time} & 
\colhead{Filter} &  
\colhead{ID} & \colhead{PI} 
} 
\startdata
    ESO057$-$SC075 & LMC & ACS/WFC & \unit[25]{s}+\unit[340]{s}\times 2 & F555W & 10595 & P. Goudfrooij\\
   & & &\unit[15]{s}+\unit[340]{s}\times 2 & F814W & & \\
	\hline
 	Hodge 11  &LMC& ACS/WFC & \unit[50]{s}\times 2 + \unit[345]{s}\times 6 + \unit[370]{s}\times 6 & F606W & 14164 & A. Sarajedini\\
   & & &\unit[70]{s}\times 2 + (\unit[345]{s}+\unit[377]{s} +\unit[410]{s})\times 6& F814W & & \\
    Hodge 11 (Ref) & & ACS/WFC & \unit[550]{s}\times4+\unit[570]{s}\times8 & F435W & 14164 & A. Sarajedini\\
   &  & &\unit[50]{s}+\unit[570]{s}\times 3 & F606W & & \\
   \hline
    IC 2146 & LMC & ACS/WFC & \unit[250]{s} & F555W & 9891 & G. Gilmore\\
   & & & \unit[170]{s} & F814W & & \\
   \hline
   NGC 1466 & LMC  & ACS/WFC & \unit[50]{s}\times 2+\unit[353]{s}\times 12 & F606W & 14164 & A. Sarajedini\\
   & & & \unit[70]{s}\times 2+(\unit[352]{s}+\unit[385]{s}+\unit[420]{s})\times 6 & F814W & & \\
    NGC 1466 (Ref) & & ACS/WFC & \unit[575]{s}\times 12 & F435W & 14164 & A. Sarajedini\\
   & & & \unit[50]{s}+\unit[566]{s}\times 3 & F606W & & \\
   \hline
    NGC 1644  &LMC& ACS/WFC & \unit[250]{s} & F555W & 9891 & G. Gilmore\\
   & & & \unit[170]{s} & F814W & & \\
   \hline
    NGC 1651 &LMC & WFC3/UVIS & \unit[120]{s}+\unit[600]{s}+\unit[720]{s} & F475W & 12257 & L. Girardi\\
   & & & \unit[30]{s}+\unit[700]{s}\times 2 & F814W & & \\
    NGC 1651 (Ref) & & ACS/WFC & \unit[500]{s}\times 2 & F475W & 12257 & L. Girardi\\
   & & & \unit[500]{s}\times 2 & F814W & & \\
   \hline
    NGC 1652 &LMC & ACS/WFC & \unit[300]{s} & F555W & 9891 & G. Gilmore\\
   & & & \unit[200]{s} & F814W & & \\
   \hline
   NGC 1718  &LMC& WFC3/UVIS & \unit[120]{s}+\unit[600]{s}+\unit[720]{s} & F475W & 12257 & L. Girardi\\
   & & & \unit[30]{s}+\unit[700]{s}\times 2 & F814W & & \\
    NGC 1718 (Ref)&  & ACS/WFC & \unit[500]{s}\times 2 & F475W & 12257 & L. Girardi\\
   & & & \unit[500]{s}\times 2 & F814W & & \\
   \hline
    NGC 1783 &LMC & ACS/WFC & \unit[40]{s}+\unit[340]{s}\times 2 & F555W & 10595 & P. Goudfrooij\\
   & & & \unit[8]{s}+\unit[340]{s}\times 2 & F814W & & \\
    NGC 1783 (Ref)&  & ACS/WFC & \unit[80]{s}+\unit[300]{s}\times 2 & F814W & 12257 & L. Girardi\\
   & & & \unit[350]{s}\times 2 & F555W & & \\
   \hline
    NGC 1806  &LMC & ACS/WFC & \unit[40]{s}+\unit[340]{s}\times 2 & F555W & 10595 & P. Goudfrooij\\
   & & & \unit[8]{s}+\unit[340]{s}\times 2 & F814W & & \\
    NGC 1806 (Ref) & & ACS/WFC & \unit[80]{s}+\unit[300]{s}+\unit[340]{s} & F814W & 12257 & L. Girardi\\
   & & & \unit[350]{s}\times 2 & F555W & & \\
   \hline
    NGC 1831  &LMC& WFC3/UVIS & \unit[100]{s}+\unit[660]{s}+\unit[720]{s} & F814W & 14688 & P. Goudfrooij\\
   & & & \unit[975]{s}\times 2+\unit[1115]{s}\times 2 & F336W & & \\
   NGC 1831 (Ref)&  & ACS/WFC & \unit[850]{s}+\unit[890]{s}+\unit[980]{s} & F435W & 14688 & P. Goudfrooij\\
   & & & \unit[600]{s}+\unit[950]{s} & F814W & & \\
   \hline
    NGC 1841  &LMC& ACS/WFC & \unit[50]{s}\times 2+\unit[353]{s}\times 12 & F606W & 14164 & A. Sarajedini\\
   & & & \unit[70]{s}\times 2+(\unit[352]{s}+\unit[385]{s}+\unit[420]{s})\times 6 & F814W & & \\
    NGC 1841 (Ref)&  & ACS/WFC &\unit[575]{s}\times 12 & F435W & 14164 & A. Sarajedini\\
   & & & \unit[50]{s}+\unit[566]{s}\times 3 & F606W & & \\
   \hline
    NGC 1846  &LMC & ACS/WFC & \unit[40]{s}+\unit[340]{s}\times 2 & F555W & 10595 & P. Goudfrooij\\
   & & & \unit[8]{s}+\unit[340]{s}\times 2 & F814W & & \\
    NGC 1846 (Ref)  &LMC & WFC3/UVIS & \unit[348]{s}\times 2 & F555W & 12326 & K. Noll\\
   & & & \unit[400]{s}\times 2 & F814W & & \\
   \hline
    NGC 1852  &LMC & ACS/WFC & \unit[330]{s} & F555W & 9891 & G. Gilmore\\
   & & & \unit[200]{s} & F814W & & \\
    NGC 1852 (Ref)&  & ACS/WFC & \unit[500]{s}\times 2 & F555W & 12257 & L. Girardi\\
   & & & \unit[350]{s}\times 2 & F814W & & \\
   \hline
    NGC 1868  &LMC& WFC3/UVIS & \unit[90]{s}+\unit[666]{s} & F814W & 14710 & A. Milone\\
   & & & \unit[830]{s}+\unit[831]{s}\times 2 & F336W & & \\
    NGC 1868 (Ref) & & ACS/WFC & \unit[42]{s}+\unit[530]{s}+\unit[827]{s} & F814W & 14710 & A. Milone\\
   & & & \unit[667]{s}+\unit[679]{s}+\unit[703]{s}+\unit[760]{s} & F475W & & \\
   \hline
    NGC 1978 & LMC & ACS/WFC & \unit[300]{s} & F555W & 9891 & G. Gilmore\\
   & & & \unit[200]{s} & F814W & & \\
    NGC 1978 (Ref)&  & ACS/WFC & \unit[500]{s}\times 2 & F555W & 12257 & L. Girardi\\
   & & & \unit[350]{s}\times 2 & F814W & & \\
   \hline
    NGC 2154  &LMC& ACS/WFC & \unit[300]{s} & F555W & 9891 & G. Gilmore\\
   & & & \unit[200]{s} & F814W & & \\
    NGC 2154 (Ref)&  & ACS/WFC & \unit[500]{s}\times 2 & F555W & 12257 & L. Girardi\\
   & & & \unit[350]{s}\times 2 & F814W & & \\
   \hline
    NGC 2173  &LMC& WFC3/UVIS & \unit[30]{s}+\unit[550]{s}+\unit[700]{s}\times 2 & F814W & 12257 & L. Girardi\\
   & & & \unit[120]{s}+\unit[700]{s}\times 2 & F475W & & \\
    NGC 2173 (Ref)&  & ACS/WFC & \unit[90]{s}+\unit[500]{s}\times 2+\unit[700]{s}\times 2 & F475W & 12257 & L. Girardi\\
   & & & \unit[10]{s}+\unit[600]{s}+\unit[690]{s}+\unit[700]{s}\times 2& F814W & & \\
   \hline
    NGC 2203  &LMC& WFC3/UVIS & \unit[30]{s}+\unit[550]{s}+\unit[700]{s}\times 2& F814W & 12257 & L. Girardi\\
   & & & \unit[120]{s}+\unit[700]{s}\times 2 & F475W & & \\
    NGC 2203 (Ref)&  & ACS/WFC & \unit[90]{s}+\unit[500]{s}\times 2+\unit[700]{s}\times 2 & F475W & 12257 & L. Girardi\\
   & & & \unit[10]{s}+\unit[550]{s}+\unit[690]{s}+\unit[713]{s}\times 2 & F814W & & \\
   \hline
    NGC 2209 &LMC & WFC3/UVIS & \unit[60]{s}+\unit[485]{s}\times 2 & F814W & 12908 & P. Goudfrooij\\
   & & & \unit[850]{s}\times 2 & F438W & & \\
    NGC 2209 (Ref)&  & ACS/WFC & \unit[361]{s}+\unit[383]{s} & F814W & 12908 & P. Goudfrooij\\
   & & & \unit[820]{s}+\unit[840]{s} & F435W & & \\
   \hline
    NGC 2213 &LMC & WFC3/UVIS & \unit[120]{s}+\unit[600]{s}+\unit[720]{s} & F475W & 12257 & L. Girardi\\
   & & & \unit[30]{s}+\unit[700]{s}\times 2 & F814W & & \\
    NGC 2213 (Ref)&  & ACS/WFC & \unit[500]{s}\times 2 & F475W & 12257 & L. Girardi\\
   & & & \unit[500]{s}\times 2 & F814W & & \\
   \hline
    NGC 2249  &LMC& WFC3/UVIS & \unit[60]{s}+\unit[425]{s}\times 2 & F814W & 12908 & P. Goudfrooij\\
   & & & \unit[825]{s}\times 2 & F438W & & \\
    NGC 2249 (Ref)&  & ACS/WFC & \unit[333]{s}+\unit[413]{s} & F814W & 12908 & P. Goudfrooij\\
   & & & \unit[795]{s}+\unit[815]{s} & F435W & & \\
   \hline
    NGC 2257  &LMC& ACS/WFC & \unit[50]{s}\times 2+\unit[353]{s}\times 6+\unit[364]{s}\times 3+\unit[525]{s}\times 2 & F606W & 14164 & A. Sarajedini\\
   & & & \unit[70]{s}\times 2+(\unit[363]{s}+\unit[400]{s})\times 6+\unit[390]{s} \times3+\unit[450]{s}\times2 & F814W & & \\
    NGC 2257 (Ref) & & ACS/WFC & \unit[575]{s}\times 12 & F435W & 14164 & A. Sarajedini\\
   & & & \unit[50]{s}+\unit[570]{s}\times 3 & F606W & & \\
   \hline
    NGC 419 &SMC & ACS/WFC & \unit[10]{s}\times2 +\unit[474]{s}\times 4 & F814W & 10396 & J. Gallagher\\
   & & & \unit[20]{s}\times 2+\unit[496]{s}\times 4 & F555W & & \\
    NGC 419 (Ref) && ACS/WFC & \unit[60]{s}+\unit[350]{s}+\unit[400]{s} & F555W & 12257 & L. Girardi\\
   & & & \unit[350]{s}+\unit[400]{s}\times 2 & F814W & & \\
   \hline
\enddata
\end{deluxetable*}

We used the {\sc dolphot2.0} package \citep{2000PASP..112.1383D} to
perform our point-spread-function photometry. {\sc dolphot2.0}
provides two useful parameters for our analysis, {\tt sharpness} and
{\tt crowding}, which we can employ to reject misclassifications and
other contaminants from the samples. A large {\tt sharpness} value may
imply the detection of a cosmic ray (positive value) or a blend,
cluster, or galaxy (negative value). The {\tt crowding} parameter
tells us how much brighter the star would have been measured had
nearby stars not been fitted simultaneously. High {\tt crowding}
values are generally a sign of poorly measured stars. We rejected all
objects with signal-to-noise ratios of less than 5 with absolute {\sc
  dolphot2.0} {\tt sharpness} parameter $\geq 0.2$ or {\sc dolphot2.0}
{\tt crowding} parameter $\geq 0.5$. These selection choices only
affect of order 3\% of the total numbers of stars brighter than
\unit[2]{mag} below the MSTO.

\subsection{Artificial Star Tests}

In order to characterize the photometric uncertainties and the
completeness levels of the {\sl HST} data, $\sim$60,000 artificial
stars were placed in both the cluster and reference fields using {\sc
  dolphot2.0}. The colors of the artificial stars were uniformly
distributed on the CMD of the cluster, while the numbers were weighted
so that larger numbers of fainter stars were generated. The stellar
locations were randomly distributed across the field. For each run, we
added 100 artificial stars so as not to significantly increase the
degree of crowding in the images. The images were then reduced using
the same measurement and selection procedures as applied to the real
data. For the BSS populations, the completenesses levels were close to
100\% in both the cluster and reference fields. The completeness
levels for stars brighter than \unit[2]{mag} below the MSTO were
typically higher than 90\%. Completeness corrections were applied as
and when necessary.

\subsection{Cluster parameters\label{sec:param}}

The cluster centers were determined based on the maximum spatial
density in the cluster field, using a two-dimensional Gaussian kernel
density estimator. We separated the cluster and reference fields into
several radial annuli and calculated their surface densities. The
number of stars in each annulus was corrected for incompleteness and
the (possibly fractional) annular area was calculated using a Monte
Carlo-based method. In essence, large numbers of randomly distributed
points were added and the number of points in a given annulus divided
by the total number of points yielded the fraction of the area with
respect to the total field of view.

While approximation of cluster profiles with a King profile
\citep{1962AJ.....67..471K} is common practice for Galactic GCs, some
evidence suggests that young clusters in weak tidal fields---such as
those in the Large Magellanic Cloud---may not (yet) be tidally
truncated. Instead, \citep[][EFF]{1989ApJ...336..734E} profiles might
be a better representation
\citep{2003MNRAS.338...85M,2003MNRAS.338..120M}. Thus we fitted the
clusters' surface densities using EFF profiles,
\begin{equation}
	\rho(r) = \rho_0\left(1+\frac{r^2}{a^2}\right)^{-\gamma/2} +
        \rho_{\mathrm{bkg}},
\end{equation}
where $\rho_0$ and$\rho_{\mathrm{bkg}}$ are the central and background
surface densities, respectively. The core radius, $a$, and the
power-law index, $c$, are linked to the King core radius,
$r_\mathrm{c}$, through
\begin{equation}
	r_\mathrm{c} = a(2^{2/\gamma}-1)^{1/2}.
\end{equation}

As regards those clusters for which we have a reference field
available, both the cluster and reference fields were used for the
surface density calculation. This approach provided excellent
constraints on the level of field-star contamination. For clusters
without a reference field, we only determined the structural
parameters for those objects exhibiting clear radial decreases towards
the science images' outer regions. The typical uncertainty in the
central coordinates of our intermediate-age clusters is smaller than
\unit[5]{arcsec}. Given their large sizes, however, the old clusters
have relatively larger uncertainties associated with their structural
parameters. The central coordinates of the old cluster NGC 1841 have
the largest uncertainties of around \unit[15]{arcsec}. Nevertheless,
the center positions we derived deviates from that of
\citet{2008MNRAS.389..678B} by less than \unit[5]{arcsec}. The same is
true for the other clusters in our sample. For most of the
intermediate-age clusters analyzed by \citet{2014ApJ...797...35G}, we
found a systematic difference in the resulting core radii,
$r_\mathrm{c}$, of less than \unit[1]{arcsec} towards smaller
values. This is likely driven by our approach to determining the
structural parameters. \citet{2014ApJ...797...35G} only considered
stars with higher completeness levels ($\geq 75$\%), which would
result in a larger core if the completeness level within the core were
relatively low. We confirmed that this systematic difference
disappears if we adopt the same method. The only exception is NGC
1783, for which \citet{2014ApJ...797...35G} reported
$r_\mathrm{c}=\unit[10.50\pm0.49]{pc}$. We note that literature
derivations of this cluster's core radius based on fitting King
profiles are discrepant. Both \citet{1991ApJS...76..185E} and
\citet{2007AJ....134.1813M} found a core radius of around \unit[5]{pc}
for NGC 1783. Compared with \citet{2003MNRAS.338...85M,
  2018MNRAS.tmp..991P}, we did not notice any significant differences
in the core radii of our old sample clusters. The clusters' structural
parameters are included in Table \ref{tab:struc}.
  
\begin{deluxetable*}{cCCCCC}[b!]
\tablecaption{Structural parameters of our MC sample clusters. \label{tab:struc}}
\tablewidth{0pt}
\tablehead{
\colhead{Cluster} &
\colhead{$\alpha_\mathrm{J2000}$ (deg)\tablenotemark{a}} &\colhead{$\delta_\mathrm{J2000}$ (deg)\tablenotemark{a}} &\colhead{$a$ (pc)} & \colhead{$\gamma$} & \colhead{$r_\mathrm{c}$ (pc)}  \\
\colhead{(1)} & \colhead{(2)} &
\colhead{(3)} & \colhead{(4)} & \colhead{(5)}& \colhead{(6)}
}
\startdata
ESO057$-$SC075 &06^\mathrm{h}13^\mathrm{m}27.30^\mathrm{s}\pm0.39^\mathrm{s}&-70\arcdeg41\arcmin43.57\arcsec\pm5.81\arcsec& 3.05\pm 0.41 & 2.07\pm 0.18 &2.97\pm 0.15 \\
Hodge 11      &06^\mathrm{h}14^\mathrm{m}22.92^\mathrm{s}\pm0.54^\mathrm{s}&-69\arcdeg50\arcmin54.90\arcsec\pm8.03\arcsec& 4.37\pm 0.41 & 2.85\pm 0.17 & 3.46\pm 0.24 \\
IC 2146       &05^\mathrm{h}37^\mathrm{m}47.43^\mathrm{s}\pm0.47^\mathrm{s}&-74\arcdeg46\arcmin58.52\arcsec\pm7.06\arcsec& 8.75\pm 1.16 & 2.89\pm 0.43 &6.86\pm 0.37 \\
NGC 1466      &03^\mathrm{h}44^\mathrm{m}32.91^\mathrm{s}\pm0.25^\mathrm{s}&-71\arcdeg40\arcmin09.85\arcsec\pm3.70\arcsec& 5.33\pm 0.78 & 4.12\pm 0.65 &3.37\pm 0.21\\
NGC 1644      &04^\mathrm{h}37^\mathrm{m}40.13^\mathrm{s}\pm0.24^\mathrm{s}&-66\arcdeg11\arcmin54.67\arcsec\pm3.57\arcsec& 2.68\pm 0.22 & 3.09\pm 0.17 &2.02\pm 0.06 \\
NGC 1651      &04^\mathrm{h}37^\mathrm{m}32.07^\mathrm{s}\pm0.33^\mathrm{s}&-70\arcdeg35\arcmin10.31\arcsec\pm4.90\arcsec& 4.94\pm 0.49 & 3.05\pm 0.23 & 3.75\pm 0.14\\
NGC 1652      &04^\mathrm{h}38^\mathrm{m}22.86^\mathrm{s}\pm0.33^\mathrm{s}&-68\arcdeg40\arcmin20.78\arcsec\pm4.96\arcsec& 2.70\pm 0.28 & 2.56\pm 0.17 & 2.29\pm 0.09 \\
NGC 1718      &04^\mathrm{h}52^\mathrm{m}25.66^\mathrm{s}\pm0.19^\mathrm{s}&-67\arcdeg03\arcmin06.91\arcsec\pm2.79\arcsec& 3.45\pm 0.32 & 3.06\pm 0.19 & 2.61\pm 0.09 \\
NGC 1783      &04^\mathrm{h}59^\mathrm{m}09.29^\mathrm{s}\pm0.40^\mathrm{s}&-65\arcdeg59\arcmin15.67\arcsec\pm6.02\arcsec& 5.63\pm 0.59 & 2.84\pm 0.20 & 4.46\pm 0.11\\
NGC 1806      &05^\mathrm{h}02^\mathrm{m}11.35^\mathrm{s}\pm0.22^\mathrm{s}&-67\arcdeg59\arcmin09.56\arcsec\pm3.26\arcsec& 4.31\pm 0.56 & 2.77\pm 0.27 & 3.47\pm 0.17 \\
NGC 1831      &05^\mathrm{h}06^\mathrm{m}16.55^\mathrm{s}\pm0.18^\mathrm{s}&-64\arcdeg55\arcmin07.62\arcsec\pm2.73\arcsec& 6.66\pm 0.78 & 4.08\pm 0.47 &4.23\pm 0.19 \\
NGC 1841     &04^\mathrm{h}45^\mathrm{m}22.84^\mathrm{s}\pm1.00^\mathrm{s}&-83\arcdeg59\arcmin47.32\arcsec\pm14.98\arcsec& 12.69\pm 1.82 & 5.12\pm 0.76 & 7.07\pm 0.39 \\
NGC 1846      &05^\mathrm{h}07^\mathrm{m}33.86^\mathrm{s}\pm0.36^\mathrm{s}&-67\arcdeg27\arcmin37.95\arcsec\pm5.40\arcsec& 5.58\pm 0.57 & 2.35\pm 0.21 & 5.01\pm 0.20\\
NGC 1852      &05^\mathrm{h}09^\mathrm{m}24.12^\mathrm{s}\pm0.34^\mathrm{s}&-67\arcdeg46\arcmin46.92\arcsec\pm5.15\arcsec& 6.42\pm 0.92 & 3.44\pm 0.54 & 4.52\pm 0.26 \\
NGC 1868      &05^\mathrm{h}14^\mathrm{m}35.74^\mathrm{s}\pm0.12^\mathrm{s}&-63\arcdeg57\arcmin14.51\arcsec\pm1.78\arcsec& 2.06\pm 0.26 &  3.15\pm 0.23 & 1.53\pm 0.07 \\
NGC 1978      &05^\mathrm{h}28^\mathrm{m}45.21^\mathrm{s}\pm0.31^\mathrm{s}&-66\arcdeg14\arcmin13.55\arcsec\pm4.65\arcsec& 5.06\pm 0.73 & 3.37\pm 0.33 &3.61\pm 0.19 \\
NGC 2154      &05^\mathrm{h}57^\mathrm{m}38.24^\mathrm{s}\pm0.21^\mathrm{s}&-67\arcdeg15\arcmin43.01\arcsec\pm3.16\arcsec& 4.37\pm 0.35 & 3.05\pm 0.20 & 3.31\pm 0.12 \\
NGC 2173      &05^\mathrm{h}57^\mathrm{m}58.35^\mathrm{s}\pm0.27^\mathrm{s}&-72\arcdeg58\arcmin43.85\arcsec\pm4.01\arcsec& 4.06\pm 0.58 & 2.96\pm 0.39 &3.14\pm 0.17 \\
NGC 2203      &06^\mathrm{h}04^\mathrm{m}43.07^\mathrm{s}\pm0.50^\mathrm{s}&-75\arcdeg26\arcmin15.98\arcsec\pm7.49\arcsec& 4.83\pm 0.46 & 2.65\pm 0.17 & 4.00\pm 0.14 \\
NGC 2209      &06^\mathrm{h}08^\mathrm{m}36.26^\mathrm{s}\pm0.76^\mathrm{s}&-73\arcdeg50\arcmin22.44\arcsec\pm10.28\arcsec& 9.91\pm 1.99 & 3.94\pm 1.05 & 6.43\pm 0.55 \\
NGC 2213      &06^\mathrm{h}10^\mathrm{m}41.95^\mathrm{s}\pm0.22^\mathrm{s}&-71\arcdeg31\arcmin46.59\arcsec\pm3.32\arcsec& 3.00\pm 0.38 & 3.31\pm 0.34 &2.16\pm 0.10  \\
NGC 2249      &06^\mathrm{h}25^\mathrm{m}49.71^\mathrm{s}\pm0.20^\mathrm{s}&-68\arcdeg55\arcmin14.66\arcsec\pm3.00\arcsec& 2.92\pm 0.24 & 3.00\pm 0.18 &2.24\pm 0.07  \\
NGC 2257     & 06^\mathrm{h}30^\mathrm{m}12.39^\mathrm{s}\pm0.46^\mathrm{s}&-64\arcdeg19\arcmin34.71\arcsec\pm6.94\arcsec&14.27\pm 1.31 & 4.60\pm 0.87 &8.46\pm 0.53 \\
NGC 419       &01^\mathrm{h}08^\mathrm{m}17.12^\mathrm{s}\pm0.34^\mathrm{s}&-72\arcdeg53\arcmin03.66\arcsec\pm5.04\arcsec& 7.02\pm 0.83 & 4.39\pm 0.44 & 4.28\pm 0.19 
\enddata
\tablenotetext{a}{The uncertainty was estimated by means of a
  bootstrap test.}  
\tablecomments{(1) Cluster name; (2,3) Central coordinates; (4) EFF
  core radius; (5) EFF concentration parameter; (6) King core radius.}
\end{deluxetable*}

We used the Padova group's PARSEC 1.2S isochrones
\citep{2012MNRAS.427..127B} to perform our CMD fits based on visual
inspection. In Table \ref{tab:iso}, we list the derived parameters:
ages, metallicties, extinction values, and distance moduli. For most
of our intermediate-age clusters, we found consistent best-fitting
metallicity values of around $Z \sim 0.007$, with the exception of NGC
419. The same cluster data were analyzed by
\citet{2016A&A...586A.148N} using identical isochrone models. These
latter authors adopted a fixed metallicity of $Z = 0.008$, which is
consistent with our results to within $\unit[1]{\sigma}$. The only
exception, NGC 419, is a star cluster in the Small Magellanic
Cloud. It has a lower metallicity compared with its similarly aged
counterparts in the Large Magellanic
Cloud. \citet{2017MNRAS.468.3150M} found best-fitting parameters for
NGC 419 of $\log (t\ \unit{yr^{-1}})=9.5$ and $Z=0.0028$, indeed
consistent with our results. As for the old clusters, we compared our
results with \citet{2017MNRAS.471.3347W} who used the same data but
applied different methods to derive the distances. The deviations for
all parameters between both sets of results are less than
$\unit[2]{\sigma}$. We also derived the clusters' central volume
  densities and their masses using the Monto Carlo method (for
  details, see the next section). Our estimates of these parameters
  are comparable to those of \citet{2003MNRAS.338...85M} and
  \citet{2005ApJS..161..304M} to within 1 to
  $\unit[2]{\sigma}$.

\begin{deluxetable*}{cCCCCCCC}[b!]
\tablecaption{Best-fitting isochrones and masses for our MC sample clusters. \label{tab:iso}}
\tablewidth{0pt}
\tablehead{
\colhead{Cluster} &
\colhead{$\log (t\ \unit{yr^{-1}})$} &   
\colhead{$Z$\tablenotemark{a}} & \colhead{$A_V$ (mag)} & \colhead{$(m-M)_0$ (mag)}
& \colhead{$m_\mathrm{ini}$ ($M_\odot$)}& \colhead{$\log \rho_0$ ($\unit{M_\odot\,pc^{-3}}$)} & \colhead{$\log M$ ($M_\odot$)} \\
\colhead{(1)} & \colhead{(2)} &
\colhead{(3)} & \colhead{(4)} & \colhead{(5)} & \colhead{(6)} & \colhead{(7)} & \colhead{(8)}}
\startdata
ESO057$-$SC075 & 9.25\pm 0.05  & 0.006 \pm 0.001  & 0.18\pm0.02  & 18.42\pm0.03 & 1.66 &1.20\pm 0.20 & 3.87\pm0.13\\
Hodge 11      & 10.09\pm 0.05 & 0.00025 \pm 0.001& 0.24\pm0.02  & 18.50\pm0.03 & 0.80&2.20\pm0.03 &5.27\pm0.03\\
IC 2146       & 9.30\pm 0.05   & 0.007 \pm 0.001  & 0.08\pm0.02  & 18.55\pm0.03 & 1.57 &0.71\pm 0.15 &4.31\pm0.09 \\
NGC 1466      & 10.10 \pm 0.05   & 0.0003 \pm 0.001 & 0.22\pm0.02  & 18.43\pm0.03 & 0.80&2.51\pm0.08 &5.20\pm0.03\\
NGC 1644      & 9.23 \pm 0.05 & 0.006  \pm 0.001 & 0.03\pm0.02  & 18.48\pm0.03 & 1.69&1.92\pm0.19 &3.91\pm0.14\\
NGC 1651      & 9.30 \pm 0.05  & 0.006 \pm 0.001  & 0.31\pm0.02  & 18.48\pm0.03 & 1.59&1.50\pm0.12&4.32\pm0.09 \\
NGC 1652      & 9.33 \pm 0.05 & 0.005 \pm 0.001  & 0.186\pm0.02 & 18.48\pm0.03 & 1.53&1.65\pm0.19 &3.91\pm0.12\\
NGC 1718      & 9.24 \pm 0.05 & 0.007 \pm 0.001  & 0.53 \pm0.02 & 18.63\pm0.03 & 1.70&2.15\pm0.09 & 4.52\pm0.07 \\
NGC 1783      & 9.25\pm 0.05  & 0.005 \pm 0.001  & 0.06 \pm0.02 & 18.46 \pm0.03 & 1.64&2.05\pm0.05 &5.00\pm0.04\\
NGC 1806      & 9.26 \pm 0.05 & 0.005 \pm 0.001  & 0.10 \pm0.02  & 18.48\pm0.03 & 1.62&2.10\pm0.06&4.71\pm0.05\\
NGC 1831      & 8.85\pm 0.05  & 0.008 \pm 0.001  & 0.12 \pm0.02 & 18.40 \pm0.03 & 2.46&1.65\pm0.12 &4.53\pm0.09\\
NGC 1841      & 10.05\pm 0.05 & 0.00025 \pm 0.001 & 0.56\pm0.02   & 18.38\pm0.03 & 0.83& 1.44\pm0.02 &5.22\pm0.03\\
NGC 1846      & 9.28\pm 0.05  & 0.006 \pm 0.001  & 0.07 \pm0.02 & 18.45 \pm0.03 & 1.62&1.60\pm0.06 &4.76\pm 0.05\\
NGC 1852      & 9.20  \pm 0.05 & 0.006 \pm 0.001  & 0.15 \pm0.02 & 18.46\pm0.03 & 1.74&1.32\pm0.11 &4.25\pm0.08\\
NGC 1868      & 9.10 \pm 0.05  & 0.006 \pm 0.001  & 0.12\pm0.02  & 18.40 \pm0.03 & 1.96&2.75\pm0.11 & 4.39\pm 0.10\\
NGC 1978      & 9.40\pm 0.05   & 0.005 \pm 0.001  & 0.10\pm0.02   & 18.55\pm0.03 & 1.45& 2.54\pm0.03 & 5.19\pm 0.03\\
NGC 2154      & 9.27 \pm 0.05 & 0.005  \pm 0.001 & 0.064\pm0.02 & 18.48\pm0.03 & 1.61& 1.67\pm0.02 & 4.32\pm0.08\\
NGC 2173      & 9.23\pm 0.05  & 0.007 \pm 0.001  & 0.24 \pm0.02 & 18.45 \pm0.03 & 1.71& 1.93\pm0.13 & 4.82\pm 0.10\\
NGC 2203      & 9.24\pm 0.05  & 0.008 \pm 0.001  & 0.19 \pm0.02 & 18.40 \pm0.03 & 1.72& 1.54\pm0.12 & 4.49\pm 0.07\\
NGC 2209      & 9.16\pm 0.05  & 0.008 \pm 0.001  & 0.20\pm0.02   & 18.45 \pm0.03 & 1.92& 0.65\pm 0.16 & 4.07\pm 0.11\\
NGC 2213      & 9.24 \pm 0.05 & 0.008  \pm 0.001 & 0.16\pm0.02  & 18.40\pm0.03 & 1.72&2.01\pm0.16 &4.10\pm 0.13\\
NGC 2249      & 9.00 \pm 0.05    & 0.006 \pm 0.001  & 0.12\pm0.02  & 18.40 \pm0.03 & 2.13&2.00\pm 0.16 & 4.09\pm 0.12\\
NGC 2257      & 10.02 \pm 0.05& 0.0005\pm 0.001  & 0.20\pm0.02   & 18.48\pm0.03 & 0.84&1.80\pm 0.03  &5.05\pm0.03\\
NGC 419       & 9.20 \pm 0.05 & 0.0025\pm 0.001  & 0.181 \pm0.02& 18.90 \pm0.03 & 1.57&2.33\pm 0.09 &5.05\pm0.04
\enddata
\tablenotetext{a}{$Z_\odot=0.0152$}
\tablecomments{(1) Cluster name; (2) Best-fitting cluster age
  (logarithmic units); (3) Metallicity; (4) Extinction; (5) Distance
  modulus; (6) Upper boundary to the initial stellar mass used to
    estimate the cluster mass; the lower boundary we adopted is
    $\unit[0.09]{\mathit{M}_\odot}$; (7) Central volume density; (8)
  Cluster mass.}
\end{deluxetable*}

\subsection{Characterization of the Stellar Populations}

We selected the BSS populations in our intermediate-age sample
clusters as follows: (a) they must be brighter than \unit[2]{mag}
below the MSTO; (b) they must also be brighter than the
\unit[250]{Myr}-old isochrone plus $\unit[3]{\sigma}$, where $\sigma$
is the photometric uncertainty; (c) they must be bluer than the
best-fitting isochrone for the cluster's bulk stellar population minus
$\unit[3]{\sigma}$. For our old clusters ($t\sim \unit[10]{Gyr}$), we
selected those stars that were bluer and brighter than the MSTO locus,
located outside the $\unit[3]{\sigma}$ envelope of the MSTO and the MS
ridge line.

\begin{figure}
\plotone{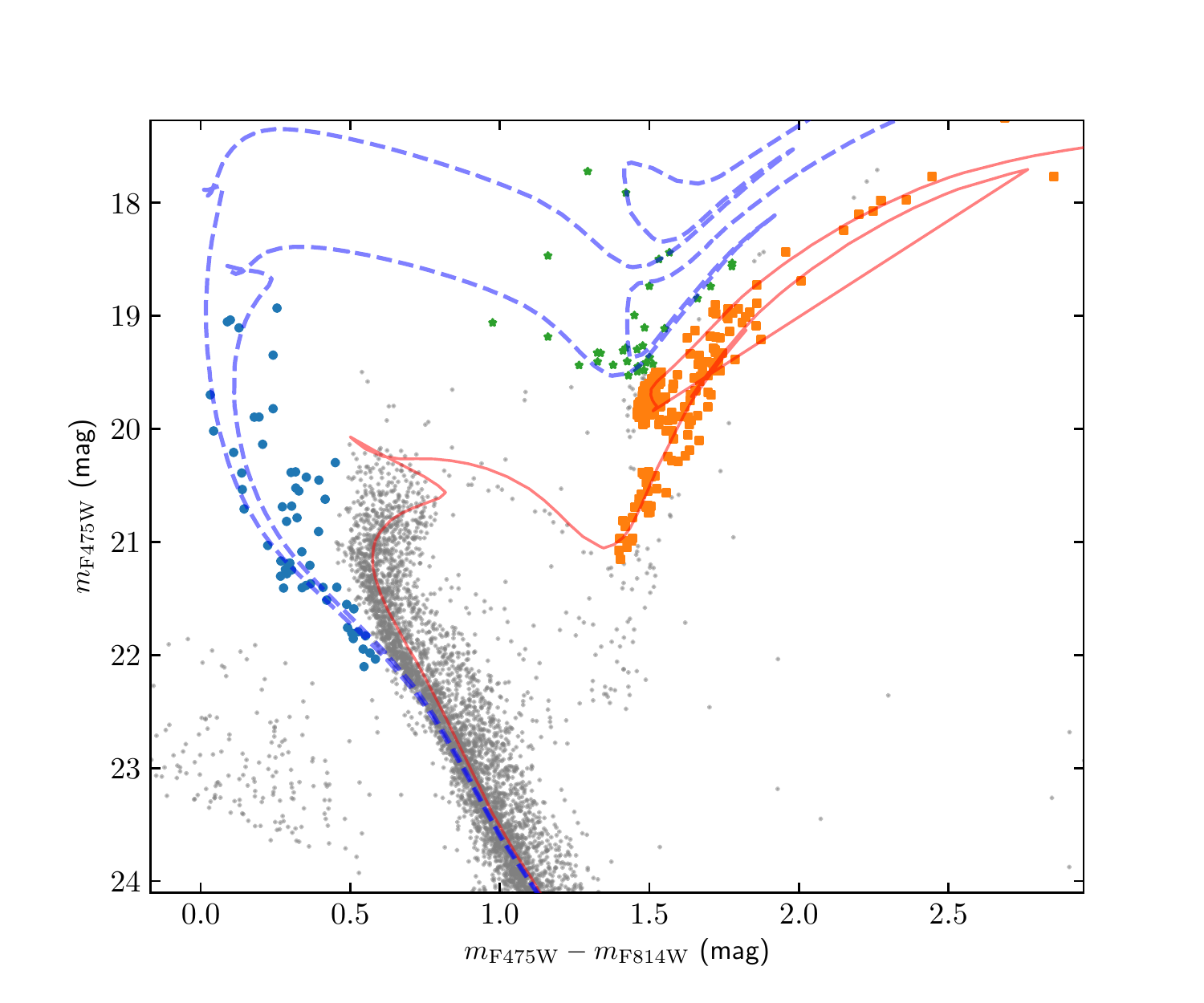}
\caption{NGC 2173 CMD. BSSs, evolved BSSs, and RGB stars are marked
  with blue circles, green pentagons, and orange squares,
  respectively. The best-fitting isochrones for the bulk stellar
  population (red solid line) and the young population (blue dashed
  line) are also shown. The CMDs pertaining to all other clusters are
  available in the Figure Set. \label{fig:cmds}}
\end{figure}

The evolved BSSs were identified using the same criteria as adopted by
Li et al. In essence, they used an isochrone with an older age to fit
the bulk stellar population, as well as two younger isochrones (see
the blue dashed lines in Fig. \ref{fig:cmds}) to roughly describe the
bottom and top boundaries of the sample of bright, evolved
stars. Next, as their evolved BSSs, they selected stars redder than
the midpoint of the subgiant branch (SGB) and located either between
these boundaries or within the $\unit[3]{\sigma}$ envelopes of these
young isochrones. Our selections of BSSs, evolved BSSs, and red
giant-branch (RGB) stars, as well as the best-fitting isochrones are
all indicated in Fig. \ref{fig:cmds}.

\figsetstart
\figsetnum{1}
\figsettitle{MC cluster CMDs}

\figsetgrpstart
\figsetgrpnum{1.11}
\figsetgrptitle{NGC 2173}
\figsetplot{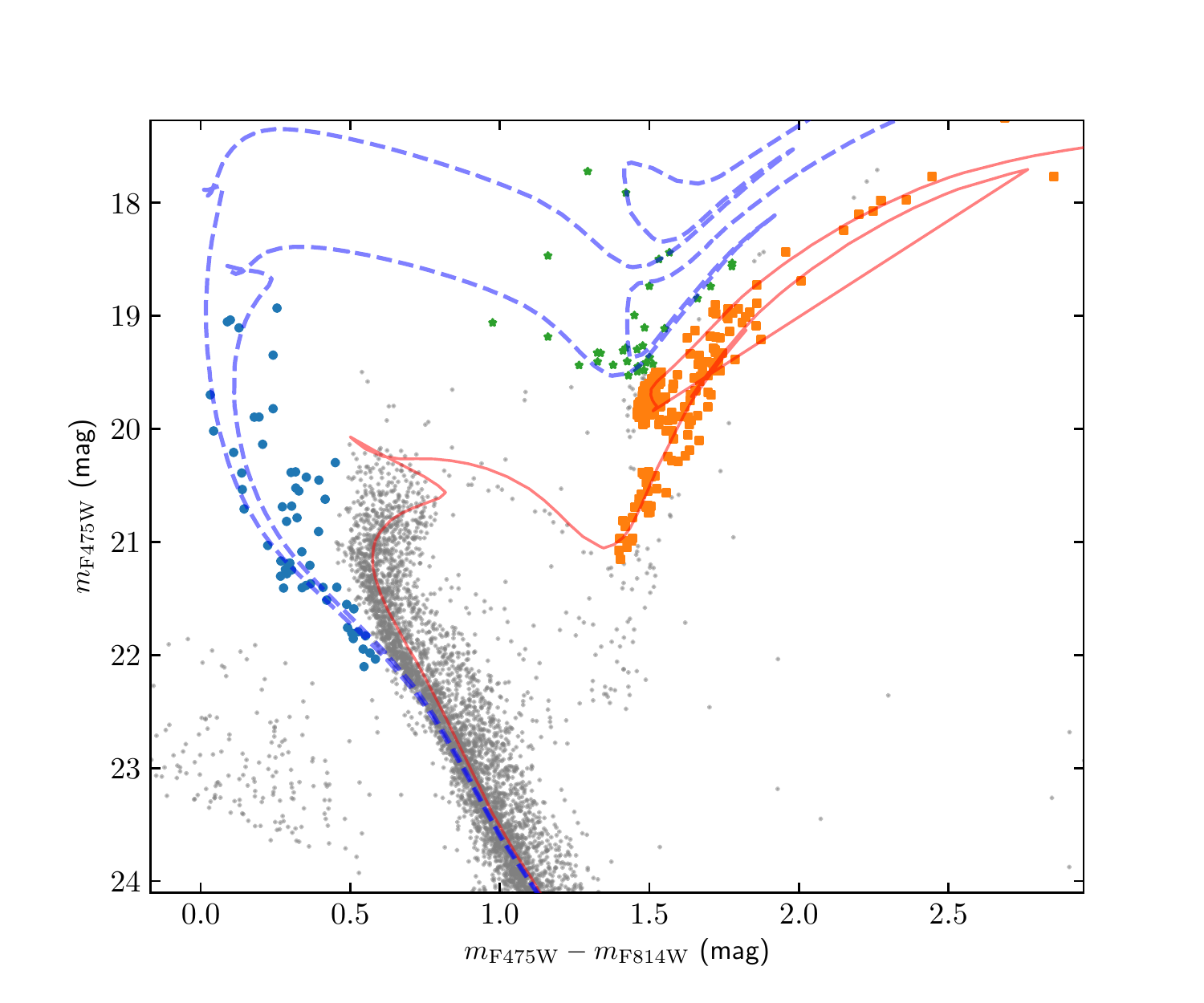}
\figsetgrpnote{NGC 2173 CMD. BSSs, evolved BSSs, and RGB stars are marked with blue circles, green pentagons, and orange squares, respectively. The best-fitting isochrones for the bulk stellar population (red solid line) and the young population (blue dashed line) are also shown. ...}
\figsetgrpend

\figsetend

In this paper, we are most interested in the total number of BSSs in
our clusters, rather than in their detailed distributions across the
CMDs. Therefore, we did not apply a field-decontamination procedure
using either stars located at the closest geometric distance in
color–-magnitude space \citep{2015A&A...575A..62N} or the number of
stars in a cell of a given color--magnitude grid
\citep{2016Natur.529..502L}. We simply calculated the total number of
objects of a given stellar population in the cluster fields and
subtracted the number of possible contaminating stars found in the
field region, after correction for the size differences of these
areas. All populations in our sample clusters returned positive values
after field decontamination, implying that the stars we selected are
most likely genuine cluster members.

\section{Analysis and Results \label{sec:result}}

One method we can use to determine the contributions from various BSS
formation channels is to derive the theoretical scaling laws and
compare these with our observational data. The number of BSSs and its
correlation with other cluster properties can potentially uncover the
dominant formation mechanism. Although it is hard to tell how an
individual BSS was formed based on a single photometric snapshot, we
can still ascertain the predominant formation channel on a statistical
basis.

We focused on the BSSs located in the cluster cores ($r \leq
r_\mathrm{c}$), where collisions are expected to be most frequent. If
single--single stellar collisions are the main trigger of BSS
formation, the number of BSSs in the cluster core should follow
$N_\mathrm{BS,c}\approx\tau_\mathrm{BS}/\tau_\mathrm{coll}$, where
$\tau_\mathrm{BS}$ is the typical BSS lifetime and
$\tau_\mathrm{coll}$ is the single star--single star collision
timescale in the cluster core. If, on the other hand, binary stars are
the primary BSS progenitors, their numbers are expected to be
proportional to both the core mass, $M_\mathrm{core}$, and the binary
fraction, $f_\mathrm{bin}$.

All previous work in this field \citep{2009Natur.457..288K,
  2011MNRAS.415.3771L, 2013MNRAS.428..897L} has thus far concluded
that scaling relations involving the core mass appear to yield the
best estimate of the BSS numbers in a cluster's core. To obtain
accurate estimates of the total stellar mass contained within the
core, we calculated membership probabilities using the structural
parameters we derived based on our EFF profile fits:
\begin{equation}
	P(r) = \frac{\rho(r)-\rho_\mathrm{bkg}}{\rho_\mathrm{bkg}}.
\end{equation}

Next, we generated a synthetic cluster using the Monto Carlo method,
adopting a Kroupa stellar initial mass function
\citep[IMF;][]{2001MNRAS.322..231K},
\begin{equation}
	\xi(m)\propto m^{-\alpha_i}
\end{equation}
where
\begin{equation}
	\begin{aligned}
		\alpha_0 =& 0.3, \quad 0.01 \leqslant m/M_\odot < 0.08 ;\\
		\alpha_1 =& 1.3, \quad 0.08 \leqslant m/M_\odot < 0.50 ;\\
		\alpha_2 =& 2.3, \quad 0.50 \leqslant m/M_\odot.\\
	\end{aligned}
\end{equation}
The best-fitting isochrone provided the upper and lower boundaries to
the initial mass of the remaining stars in the cluster, which served
as the integral range adopted for the IMF (Column 6 in Table
  \ref{tab:iso}). The initial mass, generated through Monto Carlo
sampling of the IMF, was interpolated back to the isochrone to
generate a mock cluster with the same age as the real cluster of
interest. The population of MS stars within 2 magnitudes below the
MSTO was used as control group \citep{2007ApJ...660..319X}, given that
they are not significantly affected by sampling incompleteness. The
stellar numbers in the control group (corrected for their membership
probabilities) in the real and synthetic clusters are proportional to
their masses. We then multiplied the integrated mass thus obtained for
the synthetic cluster by this ratio to estimate the total stellar mass
contained within one core radius from the real cluster's center as
well as the cluster's integrated mass (see column 8 in Table
\ref{tab:iso}).

\begin{figure}[ht!]
\plotone{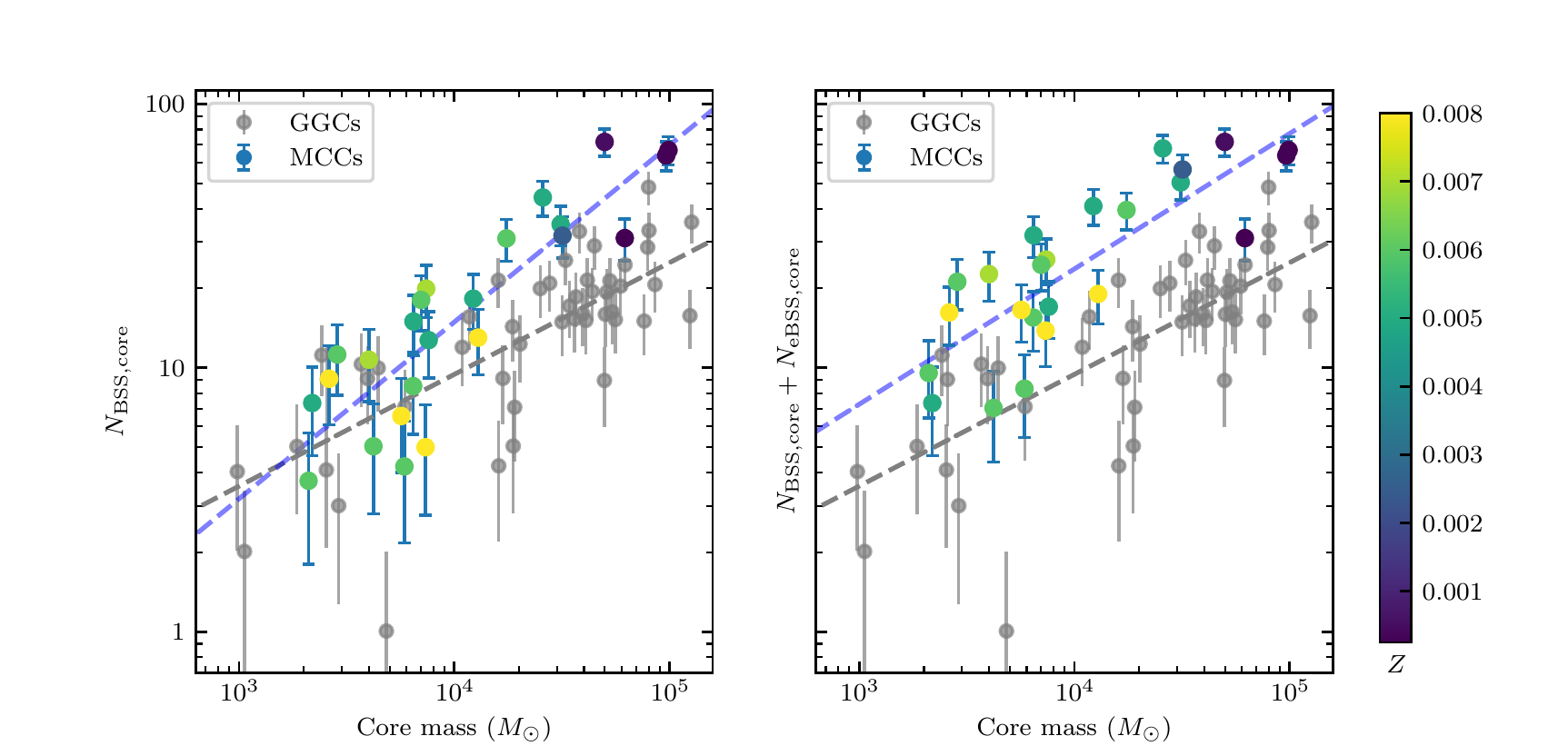}
\caption{Number of core BSSs ($N_{\mathrm{BSS, core}}$; left) and BSS
  populations including evolved BSSs ($N_{\mathrm{eBSS, core}}$;
  right) versus the core masses of our clusters. Colorful dots
  correspond to MC clusters (MCCs; colors represent the clusters'
  metallicities, $Z$), grey dots to Galactic GCs
  \citep[GGCs,][]{2009Natur.457..288K}. The error bars shown are the
  approximate 68\% Poissonian confidence intervals. The blue and grey
  dashed lines represent the best power-law fits to the
  data. \label{fig:coremass}}
\end{figure}

We found a proportionality of the type $N_\mathrm{BSS,c} \propto
{M_\mathrm{c}}^{0.66\pm 0.07}$ (see Fig. \ref{fig:coremass},
left). The uncertainty in the power-law index was estimated based on a
series of bootstrapping experiments. We created 1000 realizations and
fitted a linear relation to estimate the resulting confidence
intervals. A Spearman test indicates that the BSS numbers and cluster
core masses are strongly correlated, $\rho_\mathrm{S}=0.84$. Since the
Spearman coefficient is significantly different from either unity (by
$\unit[4.8]{\sigma}$) or zero (by $\unit[9.4]{\sigma}$), we confirm
this sublinear relation robustly. Although the resulting power-law
index is slightly larger than the value of $\sim 0.5$ found by
\citet{2009Natur.457..288K} and \citet{2013MNRAS.428..897L}, the
discrepancies are within $\unit[2]{\sigma}$.

Upon inclusion of the evolved BSSs (Fig. \ref{fig:coremass}, right),
the proportionality becomes $N_\mathrm{BSS,c} \propto
{M_\mathrm{c}}^{0.51\pm 0.07}$, with a Spearman coefficient of
$\rho_\mathrm{S}=0.80$. Now the difference in power-law indices
between our result and those published previously is less than
$\unit[1]{\sigma}$, suggesting that the BSSs in MC clusters are
characterized by a very similar correlation as their counterparts in
the Milky Way. Fig. \ref{fig:coremass} shows that the evolved BSSs are
distinguished unequivocally only in our intermediate-age clusters. In
old clusters, the locus of the evolved population is closely
coincident with that of the red clump (RC) stars. This is so, because
in optical filters the locus of the RC stars does not change
significantly for ages in excess of \unit[2]{Gyr}, thus rendering
evolved BSSs indistinguishable from RC stars associated with a
cluster's bulk stellar population
isochrone. \citet{2016ApJ...830..139P} reported the existence of such
a population in 47 Tuc using {\sl HST} data obtained in UV filters.

As we have seen, the power-law index becomes smaller when the evolved
BSSs are included in the analysis. Although the Spearman test yields a
less significant correlation for samples including evolved BSSs
compared with `clean' samples without such evolved objects, the
statistical robustness is more significant for correlations involving
cluster core masses than for any other cluster parameters
\citep{2013MNRAS.428..897L}.

Second, \citet{2009Natur.457..288K} predicted that the number of BSSs
and the number of binary stars in the core should be correlated, which
was confirmed by \citet{2013MNRAS.428..897L}. However, the large
distance to the MCs renders most methods commonly used to explore this
effect in Galactic GCs useless. \citet{2012A&A...540A..16M} measured
MS binary fractions by comparing (in a minimum-$\chi^2$ sense)
simulated CMDs characterized by a series of different binary fractions
with their observational data. Similarly, \citet{2010ApJ...724..649H}
compared synthetic CMDs characterized by different binary fractions
with observational data for the young MC cluster NGC 1818. Since we
are only able to detect high-mass-ratio binaries in the CMDs of our MC
clusters, the global binary fractions for all mass ratios rely on the
validity of the assumed mass-ratio distribution. Note, however, that
binary determinations based on CMD analysis can only be done robustly
for MS binaries, and not for binaries at the MSTO, where most
observable BSSs are expected to originate. This is a serious problem
affecting all studies attempting to deal with BSS formation based on
CMD analysis \citep{2014ApJ...797...35G}, which is further compounded
by the hotly debated physics at play in extended MSTO regions.

\begin{figure}[ht!]
\plotone{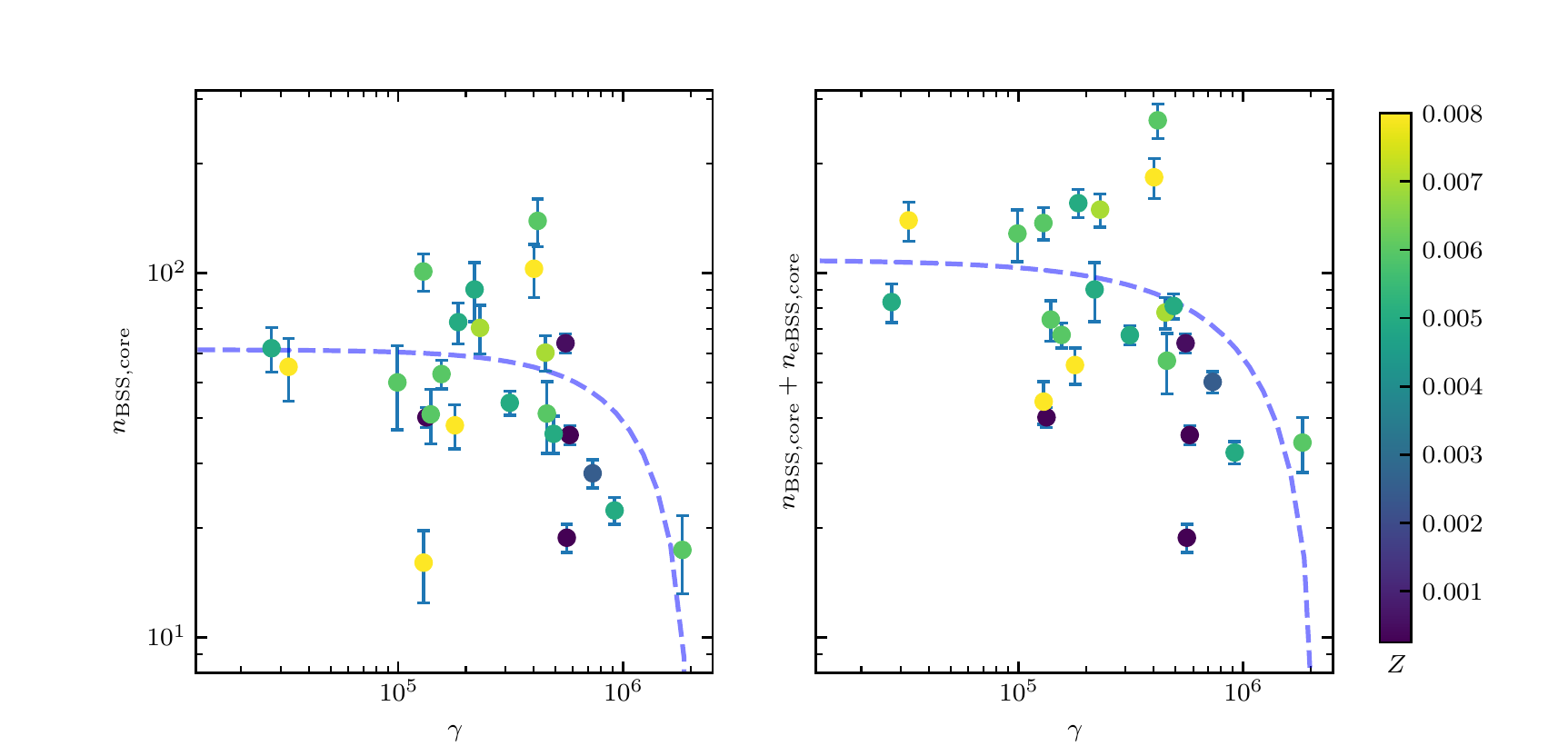}
\caption{As Fig. \ref{fig:coremass}, but for the mass-normalized
  number of BSSs in the core ($n_\mathrm{BSS,core}$) versus the
  mass-normalized collision parameter, $\gamma$. The blue dashed line
  represents the best-fitting model of the form
  $n=C+a\gamma^b$. \label{fig:collision}}
\end{figure}

In Fig. \ref{fig:collision}, we show the number of core BSSs as a
function of the collision rate, both normalized to the cluster
mass. The mass-normalized quantities are defined, respectively, as
$n_\mathrm{BSS}\equiv
N_\mathrm{BSS}/(M/\unit[10^5]{\mathit{M}_\odot})$ and $\gamma\equiv
\Gamma/(M/\unit[10^5]{\mathit{M}_\odot})$, where the collision
parameter $\Gamma$, approximately the annual stellar collision rate,
is given by
\begin{equation}
\label{eq:col}
	\Gamma =
        \left(\frac{\rho_o}{M_\odot\mathrm{pc}^{-3}}\right)^2\left(\frac{r_\mathrm{c}}{\mathrm{pc}}\right)^3\left(\frac{v_{\mathrm{c},\sigma}}{\mathrm{km\,s^{-1}}}\right)^{-1},
\end{equation}
where $\rho_0$ is the central mass volume density, $r_\mathrm{c}$ is
the core radius, and $\sigma$ is the central velocity dispersion
\citep{2006ApJ...646L.143P}. We introduced mass-normalized parameters
to alleviate the problem associated with the presence of a
$\Gamma$--mass correlation, that is, more massive clusters experience
higher integrated collision numbers, and vice versa. Following
\citet{2006ApJ...646L.143P}, we assumed both primordial and dynamical
contributions: $n=C+a\gamma^b$. Here, $C$ represents the fraction
  of the contribution that is independent of the collision properties
  and which has a primordial origin (e.g., primordial binary systems),
  while the second term on the right-hand side of the equation
  represents the fraction of the contribution that is related to
  dynamical interactions in the cluster. Fig. \ref{fig:collision}
shows a decreasing trend for larger $\gamma$, although the spread is
significant. The normalized number of BSSs may exhibit a weak (or
perhaps no) dependence on the collision rate for $\gamma \leqslant
3\times10^5$, suggesting that a collision-dominated BSS formation
mechanism is ruled out. For larger $\gamma$, collisions do not enhance
but instead suppress the formation of BSSs. This could be caused by
the effects of disruption due to close encounters on the numbers of
primordial binaries. Since BSSs formed from binaries represent the
majority of a BSS population, this difference in primordial binary
fraction could also explain the large spread. In the next section, we
will discuss the interpretation of this result in detail.

To provide a direct probe of the BSS populations in various clusters,
we introduce the specific frequency of BSSs, defined as the ratio of
the number of BSSs to that of a reference population. Numerous
approaches have been used to estimate this parameter
\citep{1995A&A...294...80F}. Here, we adopt $N_\mathrm{2,core}$ as the
reference population. $N_2$ is defined as the number of stars within a
\unit[2]{mag} interval below the MSTO. We also estimated the specific
frequency by normalizing to the number of RGB stars
(Fig. \ref{fig:cmds}). Our results were identical no matter which
reference population we used. Therefore, we adopted
$N_\mathrm{2,core}$ as our reference population, yielding in turn
$F_\mathrm{BSS}$. Moreover, we inspected the correlation, if any, of
$F_\mathrm{BSS}$ with $N_\mathrm{2,core}$ rather than with core mass
or core stellar number, because the number of BSSs is more reliable
and robust than the core mass and, by virtue of their brightness, less
affected by the level of incompleteness.
 
\begin{figure}[ht!]
\plotone{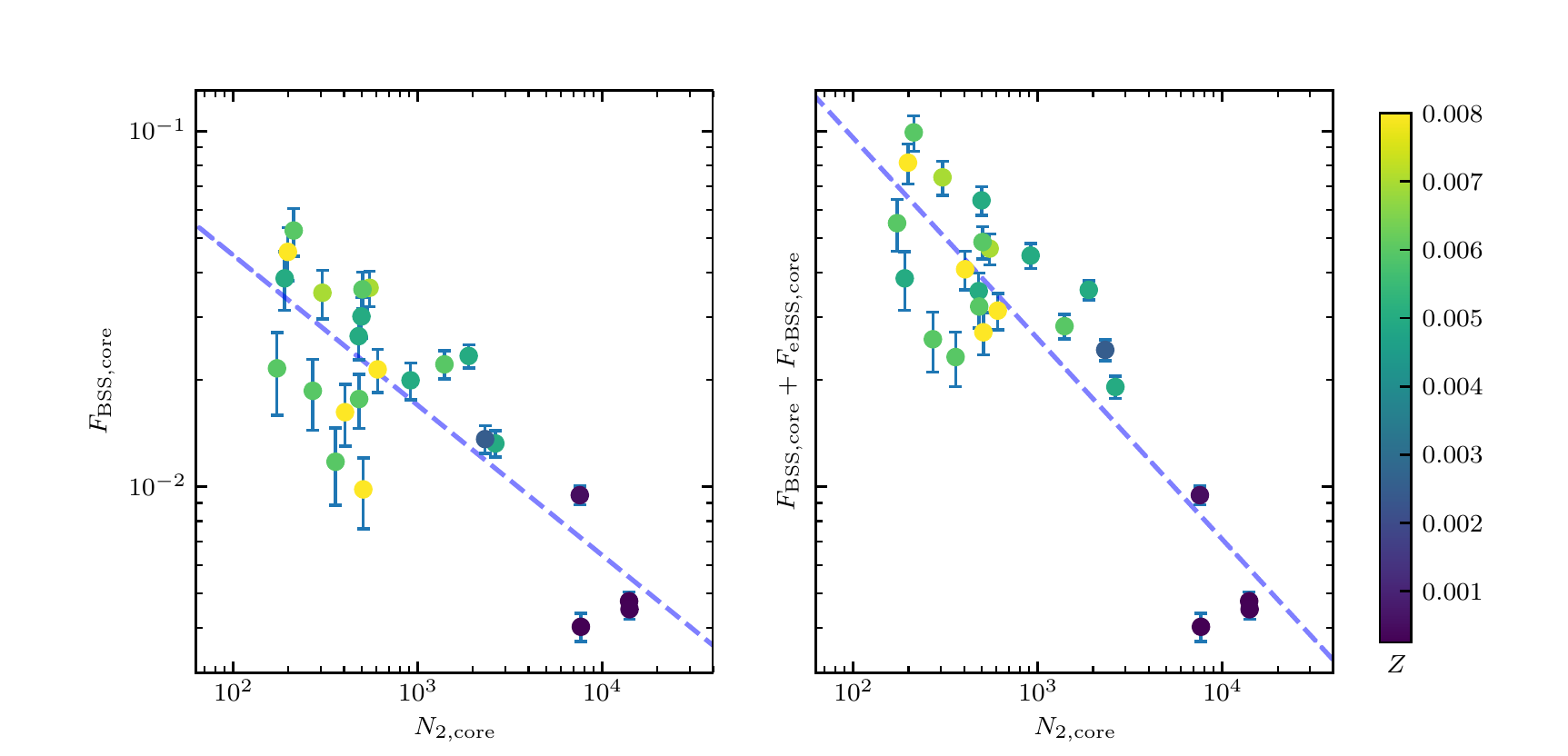}
\caption{As Fig. \ref{fig:coremass}, but for the specific frequency of
  BSSs ($F_\mathrm{BSS, core}$) versus $N_\mathrm{2,core}$. The
    specific frequencies of BSS and evolved BSS in the core are
    indicated by the subscripts `BSS' and `eBSS,'
    respectively. \label{fig:frequency}}
\end{figure}

We obtained reliable anti-correlations with $N_2$, resulting in
power-law indices of $-0.43\pm 0.06$ and $-0.56 \pm 0.07$ for BSSs and
BSSs combined with evolved BSSs, respectively. Both values differ by
$\unit[6]{\sigma}$ from zero; Spearman tests resulted in coefficients
$\rho_\mathrm{S}=-0.66$ and $-0.73$, respectively. Generally,
$N_\mathrm{2,core}$ should be proportional to a cluster's luminosity,
$M_V$, so this is also a strong indication of the presence of an
anti-correlation with total brightness. Therefore, this type of
relation, which was first reported by \citet{2004ApJ...604L.109P} for
Galactic GCs, has thus been confirmed to apply to MC clusters as well.

\section{Discussion \label{sec:discussion}}

We found a strong correlation between the number of BSSs in the core
and core mass, $N_\mathrm{BSS,c} \propto {M_\mathrm{c}}^{0.51\pm
  0.07}$, in our MC clusters, whereas we did not find any strong
dependence on the collision rate in low-collision-rate clusters. In
fact, higher collision rates suppress the formation efficiency of
BSSs. Therefore, we suspect that this may be evidence of the
predominance of the BSS binary formation channel. To quantify the
possible contributions from the binary and the collision channels, we
reproduced Fig. \ref{fig:collision} based on the simulations of
\citet{2013ApJ...777..106C}. Numerical simulations have been widely
applied to study the formation of BSSs in dense globular
clusters. They provide a unique method to explore the formation
channel of individual BSSs. \citet{2013MNRAS.429.1221H,
  2017MNRAS.466..320H} conducted Monto Carlo simulations of dense GCs
and investigated the dependence of the dominant type of BSSs on the
initial cluster conditions. \citet{2013ApJ...777..106C} found that the
collision channel increases in importance with increasing central
density. They reproduced the well-established correlation between the
number of BSSs in the cluster core with $\Gamma$. Subsequently,
\citet{2013ApJ...777..105S} carried out mock observations of their
simulated clusters and investigated correlations between their BSSs
and various cluster properties. They concluded that the strong
correlation of $N_\mathrm{BSS, c}$ with cluster core mass and the weak
dependence on $\Gamma$ found in recent observations can be
reconstructed in their models if most BSSs are created through
binary-mediated collisions. These result therefore do not support the
idea that the binary formation channel could dominate in GCs.

\begin{figure}[ht!]
\plotone{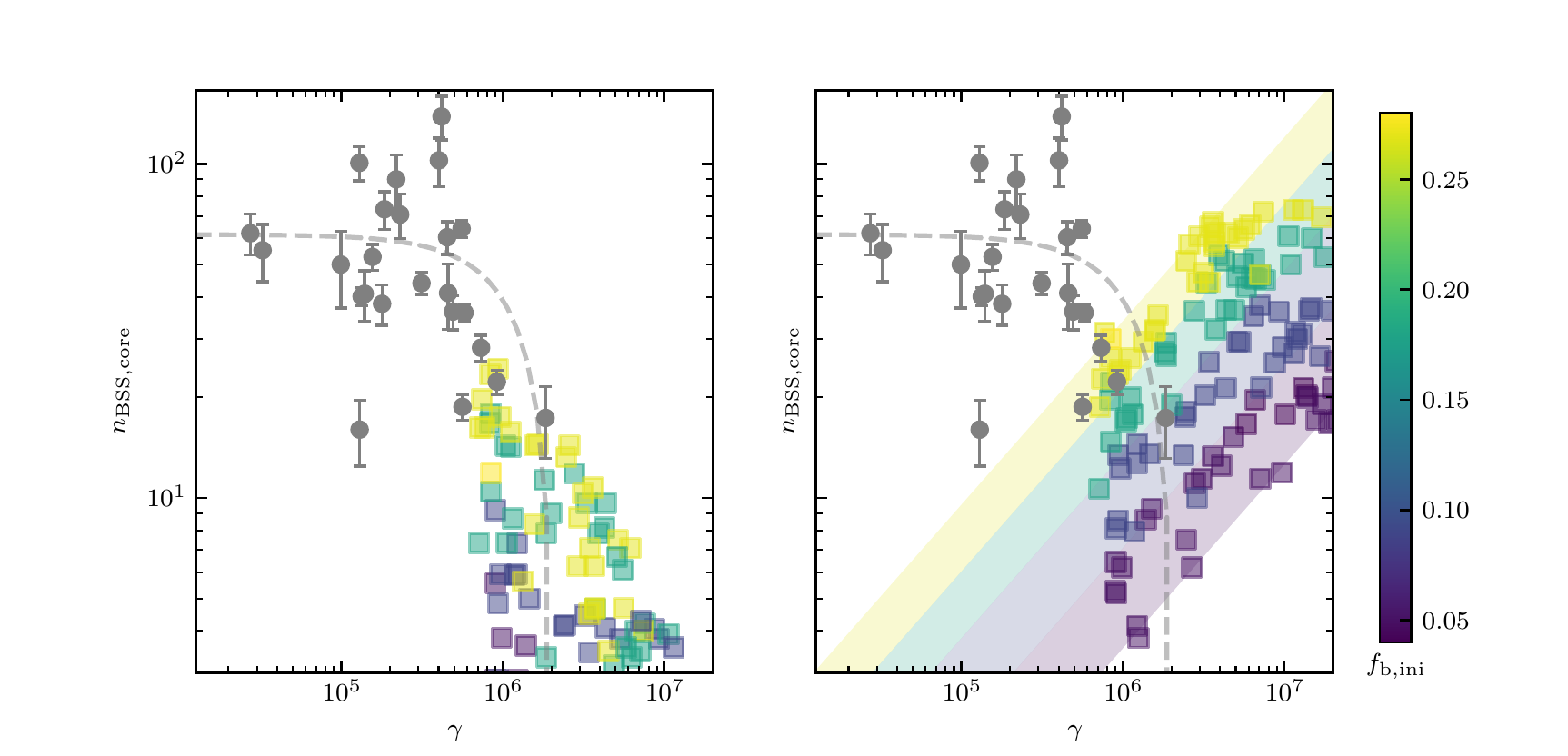}
\caption{As Fig. \ref{fig:collision}, but with the simulation of
  \citet{2013ApJ...777..106C} overplotted. The MCCs' mass-normalized
  numbers of BSSs in the core ($n_\mathrm{BSS, core}$) versus the
  mass-normalized collision parameter $\gamma$ are shown as grey dots,
  and the best-fitting models are shown as grey dashed lines (as in
  the left-hand panel of Fig. \ref{fig:collision}). BSSs formed
  through the binary and collision channels in the simulation are
  presented, respectively, in the left- and right-hand panels as
  colorful squares, with different initial binary fractions
  ($f_\mathrm{b, ini}$) indicated by different colors. The shaded
  regions encompass the possible contributions from collisions in
  lower collision-rate environments. \label{fig:mod}}
\end{figure}

Fig. \ref{fig:mod} includes the simulation result of
  \citet[][their Table 1]{2013ApJ...777..106C}. BSSs formed through
the binary and collision channels are shown as colorful squares,
respectively, in the left- and right-hand panels, with different
initial binary fractions indicated by different colors. We
extrapolated the contribution from collisions toward lower collision
rates (shadowed regions). We caution, however, that the collisional
$\Gamma$ given by Eq. \ref{eq:col} only predicts BSS formation through
single--single star collisions \citep{2013ApJ...777..106C}. A series
of simulations has proved that the probability of collisions involving
more than two stars is more significant by orders of magnitude in a
binary-mediated model \citep[e.g.][]{2004MNRAS.352....1F,
  2013ApJ...777..106C, 2013ApJ...777..105S}, since single--binary,
binary--binary, and higher-order collisions have much larger
collisional cross sections than single--single star collisions. Based
on Fig. \ref{fig:mod} we suggest that the collision parameter shows a
correlation for higher-order collisions; the actual share appears to
be regulated by the fractions of binary or triple systems.

\citet{2013ApJ...777..106C} explored a simulation grid covering star
clusters with observed central densities $\rho_{\mathrm{c, obs}}$
between $\approx 3\times10^2$ and
$\unit[4\times10^6]{\mathit{M}_\odot\,pc^{-3}}$, whereas the densities
of our MCCs (see Table \ref{tab:iso}) are generally lower than those
of Galactic GCs by an order of magnitude. Although the simulations
only marginally overlap with our observational parameter space, they
are indeed adequate to offer insights into the driving mechanism of
BSS formation in our sample clusters. The left-hand panel of
Fig. \ref{fig:mod} shows that the mass-normalized number of BSSs
generated from primordial binaries in the simulation declines toward
higher collision rates, as observed in the MCCs, while the collision
model predicts an opposite tendency. This suggests that binary
disruption may be at work. Extrapolating, we estimate that the share
contributed by the collision channel in lower collision-rate
environments is less than 20\%, demonstrating a more significant
contribution from the binary channel. Those clusters
  characterized by $7\times10^5 \leqslant \gamma\leqslant2\times 10^6$
  may be interpreted as spanning a transition phase between
  primordial-binary-dominated and collision-dominated systems, where
  $\gamma$ is, on the one hand, large enough to suppress BSS
  production via the binary channel, but on the other hand not high
  enough to form collisional BSSs efficiently.

\begin{figure}[ht!]
\plotone{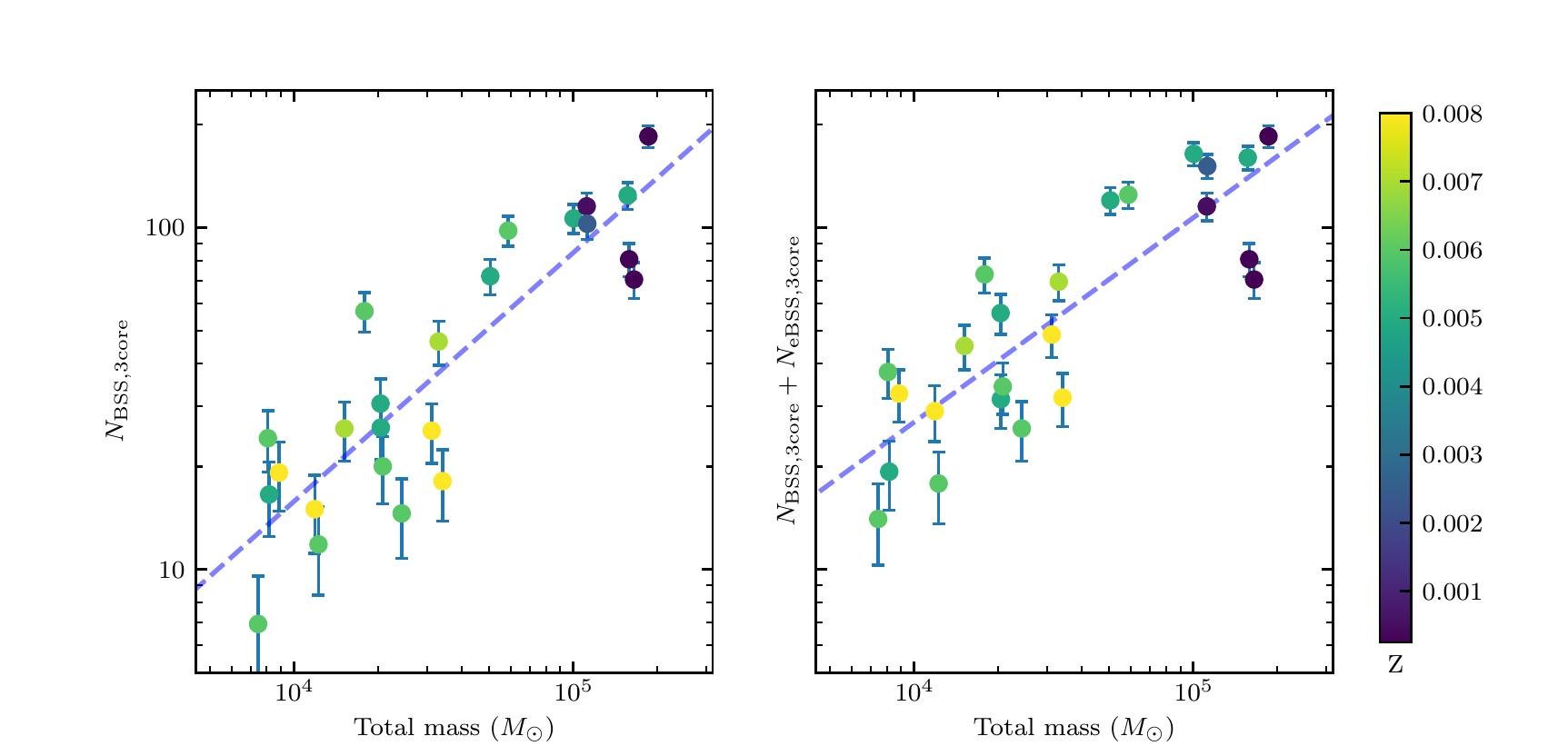}
\caption{As Fig. \ref{fig:coremass}, but for the number of BSSs within
  $\unit[3]{r_\mathrm{c}}$ versus total mass. \label{fig:3core}}
\end{figure}

We also checked the number of BSSs within a larger radial range
($\unit[3]{r_\mathrm{c}}$; see Fig. \ref{fig:3core}) and found that
the correlation with core (or total) mass was almost identical to the
earlier results (yielding power-law indices of $0.72\pm 0.08$ and
$0.60\pm 0.08$). This implies that the dominant BSS formation channel
appears universal across the cluster, which would not be expected if
the collision scenario were dominant. If the formation of BSSs in
these clusters is mainly attributed to mass transfer in binary
systems, we can assume the number of BSSs to scale with the number of
binary stars, $N_\mathrm{BSS}\propto N_\mathrm{bin} \propto
f_\mathrm{bin}M$. The proportionality $N_\mathrm{BSS,c} \propto
{M_\mathrm{c}}^{0.51\pm 0.07}$ suggests that the binary fraction is
anti-correlated with core mass, with a power-law index of $-0.5$. This
is also confirmed by Fig. \ref{fig:frequency}, where the BSS specific
frequency should scale with binary fraction in the binary-formation
scenario. Such an anti-correlation between binary fraction and core
mass has already been found for Galactic GCs
\citep{2012A&A...540A..16M}, but this is the first time that such a
relation is demonstrated for MC clusters. As a cluster evolves for
several relaxation times, dynamical mass segregation causes binaries
to sink toward the center of the cluster's gravitational potential
well, thus increasing the binary fraction, although strong dynamical
interactions can, in turn, destroy binaries. This process is at the
basis of the regulation of the binary fraction in GCs
\citep{2009ApJ...707.1533F}. Most of our sample clusters are young or
of intermediate dynamical age (their ages are comparable to their
half-mass relaxation timescales), so the effects of mass segregation
should be relatively minor. The dynamical interaction rates are also
limited because of the relatively low stellar densities. Therefore, it
is interesting that we have found that the same general
anti-correlation exists in a completely distinct environment.

\section{Conclusions \label{sec:conclusion}}

We analyzed the CMDs of 24 star clusters in the Magellanic Clouds
using multi-passband {\sl HST} images. We constructed a homogeneous
database of BSSs and their evolutionary products. We studied the
properties of the BSSs relative to the clusters' core masses, and
compared our BSS observations with the numbers expected from collision
and binary formation models. Our main results and conclusions are
summarized below:

\begin{itemize}
\item BSSs are found in all of our clusters, with BSS numbers in their
  cores ranging from five to 70. Evolved BSSs can be resolved in the
  intermediate-age clusters using optical bands, whereas they cannot
  be disentangled easily from the RC in old clusters. After field
  decontamination, we verified that these BSS populations likely
  represent genuine cluster members.

\item We confirmed that the correlation between the number of BSSs in
  a cluster's core with its core mass is valid in MC clusters, with
  power-law indices of $0.66\pm 0.07$ (only BSSs) or $0.51\pm 0.07$
  (including evolved BSSs). The discrepancy with previously published
  results is of order 1 to $\unit[2]{\sigma}$. The
  mass-normalized number of BSSs shows little dependence on the
  mass-normalized collision parameter for $\gamma \leqslant
  3\times10^5$; a decreasing trend with increasing collision parameter
  was found for larger collision rates, which is likely due to binary
  disruption. A comparison with \citet{2013ApJ...777..106C}'s
  simulations disfavors the collision channel and supports a dominant
  binary channel, defined by a contribution in excess of 80\% in lower
  collision-rate clusters ($\gamma \leqslant 3\times10^5$).

\item Since for the numbers of BSSs within a large radial range we
  still find a similar correlation with cluster mass, mass-transfer
  binaries are most likely the progenitors of the BSSs in most of our
  MC sample clusters. The anti-correlation of the BSS specific
  frequency with core mass further underscores this argument. This
  also implies an anti-correlation between the binary fraction and
  core mass.
  
\item We argue that MCCs represent a special environment for the study
  of BSS formation in clusters. Although the correlation of the number
  of BSSs with core mass in the MCCs is quite similar to that in
  Galactic GCs, the underlying formation channels might be
  different. The low stellar density (or collision rate) of MCCs
  provides a great opportunity to resolve the tension between
  observations and simulations in the interpretation of the
  correlation in Galactic GCs, which deserves further study.
\end{itemize}

\acknowledgments We thank J. Hong for enlightening discussions about
simulation aspects. C.L. acknowledges funding support from the
Macquarie Research Fellowship Scheme. R.d.G. and L.D. acknowledge
research support from the National Natural Science Foundation of China
through grants 11633005, 11473037, and U1631102. R.d.G. is grateful
for support from the National Key Research and Development Program of
China through grant 2017YFA0402702 from the Chinese Ministry of
Science and Technology (MOST). L.D. also acknowledges support from
MOST through grant 2013CB834900.
 
\software{dolphot2.0 \citep{2000PASP..112.1383D}, PARSEC
  \citep[1.2S;][]{2012MNRAS.427..127B}, 
  Astropy \citep{2013A&A...558A..33A}, 
  Matplotlib \citep{2007CSE.....9...90H}}

\end{document}